\documentclass[prd,superscriptaddress,amsfonts,amssymb,amsmath,showpacs,twocolumn]{revtex4-2}
\usepackage{bm}
\usepackage{amsfonts}
\usepackage{latexsym}
\usepackage{graphicx}
\usepackage{amsmath}
\usepackage{palatino}
\usepackage{mathpazo}
\usepackage{textcomp}
\usepackage{float}
\usepackage{booktabs}
\usepackage{dcolumn}
\usepackage{booktabs}
\usepackage{multirow}
\usepackage{hyperref}
\hypersetup{colorlinks,citecolor=blue}
\usepackage{amsmath}
\usepackage{xcolor}
\usepackage{orcidlink}
\usepackage[caption=false]{subfig}
\usepackage{commath}
\def\jnl@style{\it}
\def\aaref@jnl#1{{\jnl@style#1}}

\def\aaref@jnl#1{{\jnl@style#1}}

\def\aj{\aaref@jnl{AJ}}                   
\def\apj{\aaref@jnl{ApJ}}                 
\def\apjl{\aaref@jnl{ApJ}}                
\def\apjs{\aaref@jnl{ApJS}}               
\def\apss{\aaref@jnl{Ap\&SS}}             
\def\aap{\aaref@jnl{A\&A}}                
\def\aapr{\aaref@jnl{A\&A~Rev.}}          
\def\aaps{\aaref@jnl{A\&AS}}              
\def\mnras{\aaref@jnl{Mon.~Not.~Roy.~Astron.~Soc.}}             
\def\prd{\aaref@jnl{Phys.~Rev.~D}}        
\def\prc{\aaref@jnl{Phys.~Rev.~C}}  
\def\prl{\aaref@jnl{Phys.~Rev.~Lett.}}    
\def\qjras{\aaref@jnl{QJRAS}}             
\def\skytel{\aaref@jnl{S\&T}}             
\def\ssr{\aaref@jnl{Space~Sci.~Rev.}}     
\def\zap{\aaref@jnl{ZAp}}                 
\def\nat{\aaref@jnl{Nature}}              
\def\aplett{\aaref@jnl{Astrophys.~Lett.}} 
\def\apspr{\aaref@jnl{Astrophys.~Space~Phys.~Res.}} 
\def\physrep{\aaref@jnl{Phys.~Rep.}}      
\def\physscr{\aaref@jnl{Phys.~Scr}}       
\def\commat{\aaref@jnl{Comm.~Math.~Phys.}}              
\def\science{\aaref@jnl{Science}}               
\def\cqg{\aaref@jnl{Classical Quant.~Grav.}}            
\def\jpcs{\aaref@jnl{JPCS}}                                     
\def\ijmpd{\aaref@jnl{Int.~J.~Mod.~Phys.~D}}                    
\def\grg{\aaref@jnl{Gen.~Relat.~Gravit.}}               
\def\rpp{\aaref@jnl{Rep.~Prog.~Phys.}}          
\def\npa{\aaref@jnl{Nucl.~Phys.~A}}        
\def\lrr{\aaref@jnl{Living Rev.~Rel.}}                   
\def\jcap{\aaref@jnl{J.~Cosmology Astropart.~Phys.}}    
\def\rmp{\aaref@jnl{Rev.~Mod.~Phys.}}   
\def\epjc{\aaref@jnl{Eur.~Phys.~J.~C}}

\allowdisplaybreaks[1]

\begin{document}

\color{black}       

\title{A new $f(Q)$ cosmological model with $H(z)$ quadratic expansion}

\author{N. Myrzakulov\orcidlink{0000-0001-8691-9939}}
\email[Email: ]{nmyrzakulov@gmail.com}
\affiliation{L. N. Gumilyov Eurasian National University, Astana 010008,
Kazakhstan.}
\affiliation{Ratbay Myrzakulov Eurasian International Centre for Theoretical
Physics, Astana 010009, Kazakhstan.}

\author{M. Koussour\orcidlink{0000-0002-4188-0572}}
\email[Email: ]{pr.mouhssine@gmail.com}
\affiliation{Quantum Physics and Magnetism Team, LPMC, Faculty of Science Ben
M'sik,\\
Casablanca Hassan II University,
Morocco.} 

\author{Dhruba Jyoti Gogoi \orcidlink{0000-0002-4776-8506}}
\email[Email: ]{moloydhruba@yahoo.in}
\affiliation{Department of Physics, Dibrugarh University,
Dibrugarh 786004, Assam, India.}


\date{\today}

\begin{abstract}
We present a new $f(Q)$ cosmological model capable of reproducing late-time acceleration, i.e. $f\left( Q\right) =\lambda _{0}\left(
\lambda +Q\right) ^{n}$ by supporting certain parametrization of the Hubble parameter. By using observational data from Hubble, Pantheon, and Baryonic Acoustic Oscillations (BAO) dataset, we investigate the constraints on the proposed quadratic Hubble parameter $H(z)$. This proposal caused the Universe to transition from its decelerated phase to its accelerated phase. Further, the current constrained value of the deceleration parameter from the combined Hubble+Pantheon+BAO dataset is $q_{0}=-0.285\pm 0.021$, which indicates that the Universe is accelerating. We also analyze the evolution of energy density, pressure, and EoS parameters to infer the Universe's accelerating behavior. Finally, we use a stability analysis with linear perturbations to assure the model's stability.
\end{abstract}

\maketitle

\section{Introduction}
\label{sec1}

The General Relativity (GR) proposed by Albert Einstein in 1915, is one of the most fundamental 
and successful theories in modern physics. It has been tested and confirmed in a wide range of 
experiments and observations, from the precession of Mercury's orbit to the detection of 
gravitational waves. However, GR has limitations and is not able to explain everything in the 
Universe. One of the main limitations of GR is its inability to account for the behavior of the 
Universe on very small scales, such as those found in the quantum world. This is because GR is 
a classical theory and does not take into account the principles of quantum mechanics. Attempts 
to combine GR with quantum mechanics, such as string theory or loop quantum gravity, have been 
proposed, but these theories are still under development and lack experimental confirmation.

Furthermore, GR is unable to explain the phenomena of Dark Matter (DM) and Dark Energy (DE), 
which constitute a significant portion of the Universe. DM and DE are inferred from their 
gravitational effects on visible matter and radiation, but their nature and properties are 
still unknown. Some Modified Gravity Theories (MGT) have been proposed to explain these 
phenomena, such as the $f(R)$ gravity theory \cite{Starobinsky1,gogoi1}, but they also require 
further observational and experimental evidence. Till now, several studies in MGT regime show 
promising aspects in different fields \cite{gogoi1, gogoi2, gogoi3, gogoi4, Kou1, Kou2, Kou3, Kou4, gogoi5, Q2}.

MGTs offer compelling theoretical concepts to tackle the cosmological constant problem and 
explain the Universe's late-time acceleration. 
In an attempt to explain the Universe's late-time accelerated expansion, there has been a 
recent resurgence of $f(R)$ modified theories of gravity. In particular, studies have shown that 
the cosmic acceleration can indeed be explained through $f(R)$ gravity \cite{intro3}. Exploring 
alternative higher-order gravity theories can motivate researchers to pursue models that are 
consistent and inspired by various candidates for a fundamental theory of quantum gravity. For 
instance, string/M-theory suggests that scalar field couplings with the Gauss-Bonnet invariant 
G play a crucial role in the appearance of non-singular early-time cosmologies. These 
motivations can also be applied to the late-time Universe through an effective Gauss-Bonnet DE 
model \cite{intro4, intro4a, intro4b}. Many aspects of Gauss-Bonnet gravity have been 
extensively analyzed in the literature \cite{intro5, intro5a, intro5b}.

Simple modified gravity models, such as the $1/R$ theory, have been used to develop several DE 
models \cite{intro6, intro7}. These models show different aspects of the cosmological 
implications predicted by them. In another study, different promising $f(R)$ gravity models 
have been considered in the Palatini formalism to investigate the cosmological implications and 
comparison with observational data \cite{intro7a}. These studies show that MGT presents an 
attractive option due to its ability to provide subjective solutions to many key DE problems. 
Teleparallel gravity is an alternative theory to GR that describes gravitational interaction 
using the torsion scalar $T$ in a space-time with zero curvature 
\cite{intro8, intro9, intro10}. The theory, known as the Teleparallel Equivalent to GR (TEGR), is formulated by tetrad fields on the tangent space in the Weitzenbock 
connection, which differs from the Levi--Civita connection in GR. Hence such models may 
predict significantly different results in different domains, including cosmological 
perspectives. One advantage of working with $f(T)$ models is the ability to simplify dynamics 
and find exact solutions due to the order of field equations. Symmetric teleparallel $f(Q)$ 
gravity is another alternative theory where the covariant derivative of the metric tensor does not vanish \cite{intro11, intro12}. This theory, known as a Symmetric Teleparallel Equivalent to GR (STEGR), is based on the non-metricity scalar $Q$ and has attracted 
interest from many researchers \cite{intro15, intro16, intro17, intro18}. 
Additionally, STEGR is based on the generalization of Riemannian geometry described by Weyl 
geometry \cite{intro19}. Gravitational interaction is generally classified using three types 
of geometries: space-time curvature, torsion, and non-metricity. Theories like $f(R)$ gravity 
are basically well motivated modifications of space-time curvature, and recently, such models 
have been widely investigated in different perspectives. Due to to the dissimilarities in the 
fundamental level, non-metricity theories like $f(Q)$ gravity models may provide significantly 
different results even when one considers the model definitions analogous to $f(R)$ gravity 
models. Moreover, such modified models are capable of explaining the observational results 
without invoking the idea of DE and DM. Hence, MGT has attracted researchers' attention over 
the past few decades as it reflects current phenomena in the Universe. Therefore, 
gravitational interactions have been studied using several geometric forms 
\cite{intro20, intro21, intro22}. Recently, anisotropic nature of the spacetime has been 
investigated in the domain of $f(Q)$ gravity in Ref. \cite{Kou1}. Considering a quadratic form 
of $f(Q)$ gravity, thermodynamical aspects of Bianchi type-I Universe has been studied in Ref. 
\cite{Kou2}. Using power-law cosmology, the behaviour of the physical parameters, for example, 
energy density, pressure, EoS parameter, skewness parameter etc. has been investigated in 
$f(Q)$ gravity in Ref. \cite{Kou3}. 

In this work, we shall investigate the cosmological implications of a new $f(Q)$ gravity 
model, which is capable of reproducing late-time acceleration. We consider quadratic Hubble 
parameter and analyse the constraints on the proposed model using observational data from 
Hubble, Pantheon, and BAO. To infer the accelerating behaviour of the Universe, we thoroughly 
investigate evolution of energy density, pressure, and EoS parameters. This work will 
contribute significantly to the cosmological aspects in the domain of $f(Q)$ gravity and 
$H(z)$ quadratic expansion.

The paper is organized as follows.  In Sec. \ref{sec2}, we briefly study $f(Q)$ theory of gravity and modified Friedmann equations. In Sec. \ref{sec3}, we discuss and investigate the new cosmological $f(Q)$ model. The stability analysis for the model has been done in Sec. \ref{sec4}. Finally, in Sec. \ref{sec5}, we conclude the paper with a brief discussion.

\section{$f(Q)$ Theory and Cosmology}

\label{sec2}

Different gravity theories describe a metric-affine geometry by imposing
constraints on the affine connection \cite{Jarv}. The metric tensor $%
g_{\mu \nu }$ may be regarded as a generalization of the gravitational
potential, and it is primarily utilized to construct concepts like for
example volumes, distances, and angles, whereas the affine connection $%
\Gamma ^{\mu }{}_{\alpha \beta }$ gives parallel transport and covariant
derivatives. A basic result in differential geometry asserts that the
general affine connection can be divided into three separate components \cite%
{Hehl},%
\begin{equation}
\widetilde{\Gamma }^{\lambda }{}_{\mu \nu }=\Gamma ^{\lambda }{}_{\mu \nu
}+C^{\lambda }{}_{\mu \nu }+L^{\lambda }{}_{\mu \nu }\,.
\end{equation}

In this case, $\Gamma ^{\lambda }{}_{\mu \nu }\equiv \frac{1}{2}g^{\lambda
\beta }\left( \partial _{\mu }g_{\beta \nu }+\partial _{\nu }g_{\beta \mu
}-\partial _{\beta }g_{\mu \nu }\right) $ represents the Levi-Civita
connection of the metric tensor $g_{\mu \nu }$, $C^{\lambda }{}_{\mu \nu
}\equiv \frac{1}{2}T^{\lambda }{}_{\mu \nu }+T_{(\mu }{}^{\lambda }{}_{\nu
)} $ represents the contortion, with the torsion tensor specified as $%
T^{\lambda }{}_{\mu \nu }\equiv 2\Gamma ^{\lambda }{}_{[\mu \nu ]}$, and $%
L^{\lambda }{}_{\mu \nu }$ represents the disformation, 
\begin{equation}
L^{\lambda }{}_{\mu \nu }\equiv \frac{1}{2}g^{\lambda \beta }\left( -Q_{\mu
\beta \nu }-Q_{\nu \beta \mu }+Q_{\beta \mu \nu }\right) \,,
\end{equation}%
which is expressed in terms of the non-metricity tensor, $Q_{\alpha \mu \nu
}\equiv \nabla _{\alpha }g_{\mu \nu }$. In this paper, we will concentrate
on a torsion and curvature free geometry described only by non-metricity $%
Q_{\alpha \mu \nu }$. The$f(Q)$ theory is a STG modification in which a
matter Lagrangian $\mathcal{L}_{m}$ may be represented by an arbitrary
function of $Q$, where $Q$ is the non-metricity scalar. The total
gravitational action of $f(Q)$ gravity becomes \cite{J1, J2}, 
\begin{equation}
S=\int \frac{1}{2}f(Q)\sqrt{-g}~d^{4}x+\mathcal{L}_{m}\sqrt{-g}~d^{4}x,
\label{qqm}
\end{equation}%
where we set $8\pi G=1$. Moreover, $g$ and $\mathcal{L}_{m}$\ indicate the
metric determinant and the matter lagrangian density respectively. Let us
note that action (\ref{qqm}) has been proved to be equivalent to GR in flat
space for $f(Q)=-Q$ \cite{J2}. As a result, any change from GR may be converted into $%
f(Q)$.

The tensor of non-metricity is the basic concept in this family of theories,
defined as,%
\begin{equation}
Q_{\alpha \mu \nu }=\nabla _{\alpha }g_{\mu \nu }\,,
\end{equation}%
and its two independent traces are as follows: 
\begin{equation}
Q_{\alpha }=Q_{\alpha }{}^{\mu }{}_{\mu }\,,\quad \tilde{Q}_{\alpha }=Q^{\mu
}{}_{\alpha \mu }\,.
\end{equation}

Also, it is important to introduce the concept of superpotential, 
\begin{eqnarray}
4P^{\alpha }{}_{\mu \nu } &=&-Q^{\alpha }{}_{\mu \nu }+2Q_{(\mu %
\phantom{\alpha}\nu )}^{\phantom{\mu}\alpha }-Q^{\alpha }g_{\mu \nu }  \notag
\\
&&-\tilde{Q}^{\alpha }g_{\mu \nu }-\delta _{(\mu }^{\alpha }Q_{\nu )}\,.
\label{super}
\end{eqnarray}

Here, $Q=-Q_{\alpha \mu \nu }P^{\alpha \mu \nu }$ can be easily verified,
utilizing our sign conventions that are similar to those in Ref.. The tensor
of energy-momentum is given by 
\begin{equation}
T_{\mu \nu }=-\frac{2}{\sqrt{-g}}\frac{\delta \sqrt{-g}\,{\mathcal{L}}_{m}}{%
\delta g^{\mu \nu }}\,,  \label{emt}
\end{equation}%
and for notational simplicity, we propose the following definition $f_{Q}=%
\frac{df}{dQ}$. By varying the action (\ref{qqm}) with regard to the metric
tensor components yield 
\begin{eqnarray}
&&\frac{2}{\sqrt{-g}}\nabla _{\alpha }\left( \sqrt{-g}f_{Q}P^{\alpha
}{}_{\mu \nu }\right) +\frac{1}{2}g_{\mu \nu }f  \notag \\
&&+f_{Q}\left( P_{\mu \alpha \beta }Q_{\nu }{}^{\alpha \beta }-2Q_{\alpha
\beta \mu }P^{\alpha \beta }{}_{\nu }\right) =-T_{\mu \nu }\,,  \label{efe}
\end{eqnarray}%
and varying (\ref{qqm}) with regard to the connection, one obtains 
\begin{equation}
\nabla _{\mu }\nabla _{\nu }\left( \sqrt{-g}f_{Q}P^{\mu \nu }{}_{\alpha
}\right) =0\,.  \label{cfe}
\end{equation}

In this paper, we consider the Friedmann-Lema\^{\i}tre-Robertson-Walker
(FLRW) cosmology, which is a frequently used solution to Einstein's GR
equations that explains the large-scale structure of the Universe. The FLRW
metric presupposes that the Universe is homogeneous and isotropic, which
means that it has the same properties at every point and in every direction,
respectively. The flat FLRW metric is given by%
\begin{equation}
ds^{2}=-dt^{2}+a^{2}(t)\left[ dr^{2}+r^{2}\left( d\theta ^{2}+\sin
^{2}\theta d\phi ^{2}\right) \right] ,  \label{FLRW}
\end{equation}%
where $a(t)$ represents the scale factor, $t$ represents cosmic time, and $%
\left( t,r,\theta ,\phi \right) $ represents spherical coordinates. The
scale factor specifies the current size of the Universe and is used to
explain its expansion or contraction. The non-metricity scalar can be
obtained as $Q=6H^{2}$, where $H=\frac{\overset{.}{a}}{a}$ is the Hubble
parameter, which measures the rate of expansion of the Universe.

Further, we consider the distribution of matter and energy in the Universe
to be of the perfect fluid type. So, the mathematical representation of a
perfect fluid is given by the following energy-momentum tensor, $T_{\mu \nu
}=\left( \rho +p\right) u_{\mu }u_{\nu }+pg_{\mu \nu }$, where $u_{\mu }$ is
the 4-velocity satisfying the condition $u_{\mu }u^{\mu }=-1$, $\rho $ and $%
p $ are the energy density and pressure of a perfect fluid respectively. The
fluid is assumed to be isotropic in this form of the energy-momentum tensor,
which means that its properties are the same in all directions. It also
implies the fluid is at thermodynamic equilibrium, which means it may be
characterized by a barotropic equation of state (EoS) of the form $p=p\left(
\rho \right) $.

The modified Friedmann equations that describe the expansion of the Universe
are derived from field equations of $f(Q)$ gravity and are given by: 
\begin{equation}
3H^{2}=\frac{1}{2f_{Q}}\left( -\rho +\frac{f}{2}\right) ,  \label{F1}
\end{equation}%
\begin{equation}
\dot{H}+3H^{2}+\frac{\dot{f}_{Q}}{f_{Q}}H=\frac{1}{2f_{Q}}\left( p+\frac{f}{2%
}\right) ,  \label{F2}
\end{equation}%
where the dot $(\overset{.}{})$ represent derivatives with regard to cosmic
time $t$. The first equation, called the Friedmann equation, connects the
Universe's expansion rate to its energy density. The second equation, called
the acceleration equation, shows how the rate of expansion of the Universe
varies through time.

From Eqs. (\ref{F1}) and (\ref{F2}), we get the expressions of density $\rho 
$, isotropic pressure $p$ and EoS parameter $\omega $ respectively as, 
\begin{equation}
\rho =\frac{f}{2}-6H^{2}f_{Q},  \label{rho}
\end{equation}%
\begin{equation}
p=\left( \overset{.}{H}+\frac{\overset{.}{f_{Q}}}{f_{Q}}H\right) \left(
2f_{Q}\right) -(\frac{f}{2}-6H^{2}f_{Q}),  \label{p}
\end{equation}
\begin{equation}
\omega =\frac{p}{\rho }=-1+\frac{\left( \overset{.}{H}+\frac{\overset{.}{%
f_{Q}}}{f_{Q}}H\right) \left( 2f_{Q}\right) }{\left( \frac{f}{2}%
-6H^{2}f_{Q}\right) }.  \label{omega}
\end{equation}

Moreover, the gravitational action (\ref{qqm}) is reduced to the standard
Hilbert-Einstein form in the limiting case $f\left( Q\right) =-Q$. For this
scenario, we regain the so-called STEGR \cite{Lazkoz}, and Eqs. (\ref{rho})
and (\ref{p}) reduce to the standard Friedmann equations of GR, $3H^{2}=\rho 
$, and $2\dot{H}+3H^{2}=-p$, respectively. The above-mentioned modified
Friedmann equations system consists of only two independent equations with
four unknowns $\rho $, $p$, $H$, and $f$. We require two additional
constraint equations to fully solve the system and examine the time
evolution of the energy density, isotropic pressure, and EoS parameter. In
the next section, we will solve these equations using certain assumptions
derived from the literature.

\section{New Cosmological $f(Q)$ model}

\label{sec3}

With a functional form of $f\left( Q\right) $,\ the equations system (\ref%
{F1}) and (\ref{F2}) is reduced to two equations and three unknowns, with
one extra constraint to be added. In this work, a new $f\left( Q\right) $
gravity model has been discussed. The objective is to investigate the
evolution of the cosmological parameters, namely the deceleration parameter $%
q$ and the EoS parameter. Motivated by the work of Mukherjee and Banerjee \cite{Mukherjee} in 
$f(R)$ gravity, we assume the new functional form of $f\left( Q\right) $
symmetric teleparallel gravity to be $f\left( Q\right) =\lambda _{0}\left(
\lambda +Q\right) ^{n}$, where $\lambda _{0}$, $\lambda $ and $n$\ are
constants and represent the model parameters. Since $f\left( Q\right) $ must
have the dimension of $Q$, the constant $\lambda _{0}$ is present to address
the dimension. Also, due to the positive energy density, the value of the
constant $\lambda _{0}$ is assumed to be $-1$ in all future analyses. Now,
by using this form in Eqs. (\ref{rho})-(\ref{omega}), the expressions of
energy density $\rho $, isotropic pressure $p$ and EoS parameter $\omega $
reads respectively as,%
\begin{equation}
\rho =\frac{\lambda _{0}}{2}\chi ^{n-1}\left[ H^{2}(6-12n)+\lambda \right] ,
\label{rho1}
\end{equation}%
\begin{equation}
p=\frac{\lambda _{0}}{2}\chi ^{n}\left[ \frac{4n}{\chi }\left( \frac{12%
\overset{.}{H}H^{2}(n-1)}{\chi }+\overset{.}{H}+3H^{2}\right) -1\right] ,
\label{p1}
\end{equation}%
\begin{equation}
\omega =-1-\frac{4\overset{.}{H}(n-1)}{\chi }+\frac{4\overset{.}{H}(2n-1)}{%
H^{2}(6-12n)+\lambda }.  \label{omega1}
\end{equation}%
where $\chi =6H^{2}+\lambda $.

\textbf{For $\lambda_{0} = -1$, we observe from Eq. (\ref{rho1}) that in order to ensure a positive energy density, it is necessary to consider $n > 1/2$ for all values of $\lambda$. Thus, the range of $n$ that is used in our analysis, ensures a positive energy density}. Now, we require one further ansatz to examine the evolution of the cosmological
parameters. In the literature, there are various justifications for using
these equations \cite{H1, H2, DP1, DP2, DP3, DP4, DP5, DP6, DP7}. The technique is well famous as the model-independent way
to study cosmological models because it generally considers parametrizations
of any kinematic parameters such as the Hubble parameter, deceleration
parameter, and jerk parameter and gives the necessary extra equation. The
parameterization of the Hubble parameter is crucial in establishing the
nature of the Universe's expansion rate. These approaches have been widely
explored in the literature to characterize difficulties with cosmological
inquiries, such as the problem of all-time decelerating expansion, the
initial singularity problem, the horizon problem, and Hubble tension.
Generally, this approach has both advantages and disadvantages. One
advantage is that it is not affected by the Universe's matter and energy
content. One problem of this approach is that it does not explain why the
expansion is accelerating \cite{Pacif1, Pacif2}. The normalized Hubble parameter is considered to
be parametrized in this study. In terms of redshift $z$, we assume the
commonly used quadratic expansion parametric version of the normalized
Hubble parameter $\frac{H\left( z\right) }{H_{0}}=E\left( z\right) =1+\alpha
z+\beta z^{2}$, where $\alpha $ and $\beta $ are free parameters \cite{qua}. In this
case, the transition redshift $z_{tr}$ can be produced opposite to the
linear expansion i.e. $\beta =0$.

Using this ansatz, the first derivative with respect to the cosmic time of
the Hubble parameter can be written in terms of redshift as,
\begin{eqnarray}
\overset{.}{H} &=&-\left( 1+z\right) H\left( z\right) \frac{dH\left(
z\right) }{dz},  \notag \\
&=&-H_{0}^{2}(z+1)(\alpha +2\beta z)\left[ z(\alpha +\beta z)+1\right] ,
\end{eqnarray}%
where $H_{0}=H\left( 0\right) $ is the present value of the Hubble parameter.

Also, the general expression for the deceleration parameter $q(z)$ is given
by%
\begin{equation}
q\left( z\right) =-\frac{\overset{..}{a}}{aH^{2}}=-1+\frac{\left( 1+z\right) }{%
H\left( z\right) }\frac{dH\left( z\right) }{dz}.  \label{qz}
\end{equation}

Again, by using the previous ansatz and Eq. (\ref{qz}), we obtain%
\begin{equation}
q\left( z\right) =-1+\frac{(z+1)(\alpha +2\beta z)}{\beta z^{2}+\alpha z+1}.
\label{qz1}
\end{equation}

Our main objective is to test this scenario using current cosmological
observations. According to the previous discussion, our model contains two
main parameters ($\alpha $ and $\beta $) in addition to $H_{0}$. Because the
model parameters $\lambda $ and $n$ are not explicitly contained in the
Hubble parameter expression (we used the model-independent method), we try
to fix them in order to investigate the evolution of energy density,
isotropic pressure, and EoS parameters. We took the values $\lambda =12$ and
various ranges of $n=0.7,0.8,0.9$ into account. These values are chosen to be consistent with the basic observational requirement. In the next section, we will
try to constrain the model parameter $H_{0}$, $\alpha $, and $\beta $ values
using the most recent cosmological dataset.

\subsection{Observational data and fitting method}

Next, the various observational dataset can be used to constrain the
parameters $H_{0}$, $\alpha $, and $\beta $. We utilize the standard
Bayesian approach to investigate the observational data, and a Markov Chain
Monte Carlo (MCMC) method to get the posterior distributions of the
parameters. We also utilize the \textit{emcee} package for MCMC analysis 
\cite{Mackey/2013}. In this study, we employ three dataset: Hubble
dataset with 57 data points, Supernovae (SNe) dataset with 1048 data
points from Pantheon samples compilation dataset, and Baryonic Acoustic
Oscillation (BAO) dataset with six data points. Furthermore, the
probability function ${\mathcal{L}}\propto e^{-\frac{\chi ^{2}}{2}}$ gives
the best-fit values for the parameters using the pseudo-chi-squared function 
$\chi ^{2}$. Further, for all dataset, we employed 100 walkers and 500
steps to get findings, and we used the following prior for our study:
\begin{table}[h]
\centering 
\begin{tabular}{cc}
\hline\hline
Parameter & prior \\ \hline
$H_{0}$ & (60,80) \\ 
$\alpha $ & (0,1) \\ 
$\beta $ & (0,1) \\ \hline\hline
\end{tabular}%
\caption{Priors for parameter model $H_{0}$, $\protect\alpha $, and $\protect%
\beta $.}
\label{tab1}
\end{table}

\begin{itemize}
\item \textbf{Hubble dataset:} The significance of the Hubble parameter
study lies in the investigation of the expanding Universe. Furthermore, this
may be represented in terms of the redshift parameter $z$, which is relevant
in a variety of situations. We can derive the Hubble parameter's value at
certain redshifts. One of the more successful ways in this regard is
determining its value from line-of-sight BAO dataset. The differential age
technique is another popular method for determining $H(z)$. The 31 Hubble
points acquired using differential age (DA) technique and 26 points
collected from various methods including BAO, give 57 points of the Hubble
dataset in the redshift range $0.07<z<2.41$ {\cite{Sharov/2018}}. As
previously stated, we now use the pseudo chi-square function to estimate the
values of the parameters $H_{0}$, $\alpha $, and $\beta $. It is given for
the Hubble dataset by, 
\begin{equation}
\chi _{Hubble}^{2}(H_{0},\alpha ,\beta )=\sum_{i=1}^{57}\frac{%
[H_{i}^{th}(H_{0},\alpha ,\beta ,z_{i})-H_{i}^{obs}(z_{i})]^{2}}{\sigma
_{Hubble}^{2}(z_{i})},
\end{equation}%
where $H_{i}^{obs}$ is the observed value, $H_{i}^{th}$ is the Hubble's
theoretical value, and $\sigma _{z_{i}}$ is the standard error in the
observed value.

\item \textbf{Pantheon dataset:} The SNe is crucial in describing the
expanding Universe. Furthermore, spectroscopically acquired SNe data from
surveys such as the SuperNova Legacy Survey (SNLS), Sloan Digital Sky Survey
(SDSS), Hubble Space Telescope (HST) survey, Panoramic Survey Telescope and
Rapid Response System (Pan-STARRS1) give strong evidence in this direction.
The Pantheon dataset, the most current SNe data sample, contains 1048
magnitudes for the distance modulus measured throughout the range of $%
0.01<z<2.3$ for the redshift $z$ \cite{Scolnic/2018}. For the Pantheon
dataset, the pseudo chi-square function is given by, 
\begin{equation}
\chi _{SNe}^{2}(H_{0},\alpha ,\beta )=\sum_{i,j=1}^{1048}\Delta
\mu _{i}\left( C_{SNe}^{-1}\right) _{ij}\Delta \mu _{j},
\label{4b}
\end{equation}%
where $C_{SNe}$ denotes the covariance matrix \cite{Scolnic/2018}, and $%
\Delta \mu _{i}=\mu ^{th}(z_{i},\theta )-\mu _{i}^{obs}(z_{i})$
denotes the difference between the measured distance modulus value acquired
from cosmic measurements and its theoretical values estimated from the model
with the specified parameter space $\left( H_{0},\alpha ,\beta \right) $.
The theoretical and observed distance modulus are denoted by $\mu _{i}^{th}$
and $\mu _{i}^{obs}$, respectively. The theoretical distance modulus is $\mu
_{i}^{th}(z)=m-M=5LogD_{l}(z)$, in which $m$ and $M$ are the apparent and
absolute magnitudes of a standard candle, respectively. Also, the luminosity
distance is $D_{l}(z)=(1+z)\int_{0}^{z}\frac{dy}{H(y,H_{0},\alpha ,\beta)}$.

\item \textbf{BAO dataset:} The BAO distance dataset, which comprises
the 6dFGS, SDSS, and WiggleZ surveys, contains BAO values at six distinct
redshifts, as shown in Tab. \ref{tab2}. Also, the characteristic scale of
BAO is governed by the sound horizon $r_{s}$ at the epoch of photon
decoupling $z_{\ast }$, which is determined by the following relation: 
\newline
\begin{equation}
r_{s}(z_{\ast })=\frac{c}{\sqrt{3}}\int_{0}^{\frac{1}{1+z_{\ast }}}\frac{da}{%
a^{2}H(a)\sqrt{1+(3\Omega _{b0}/4\Omega _{\gamma 0})a}},
\end{equation}%
where, $\Omega _{b0}$ and $\Omega _{\gamma 0}$ are the current density of
baryons and photons, respectively. In BAO measurements, the following
relationships are employed,%
\begin{equation}
\triangle \theta =\frac{r_{s}}{d_{A}(z)},
\end{equation}%
\begin{equation}
d_{A}(z)=\int_{0}^{z}\frac{dy}{H(y)},  \label{4d}
\end{equation}
\end{itemize}
\begin{equation}
\triangle z=H(z)r_{s},
\end{equation}%
where $\triangle \theta $ denotes the observed angular separation, $%
\triangle z$ denotes the measured redshift separation of the BAO feature in
the two-point correlation function of the galaxy distribution on the sky
along the line of sight. In this study, BAO dataset of six points for $%
d_{A}(z_{\ast })/D_{V}(z_{BAO})$ are collected from the Refs. \cite{BAO1,
BAO2, BAO3, BAO4, BAO5, BAO6}, where $z_{\ast }\approx 1091$ is the redshift
at the epoch of photon decoupling and $d_{A}(z)$ is the co-moving angular
diameter distance combined with the dilation scale $D_{V}(z)=\left[
d_{A}(z)^{2}z/H(z)\right] ^{1/3}$. For the BAO dataset, the pseudo
chi-square function is assumed to be $\chi _{BAO}^{2}=X^{T}C_{BAO}^{-1}X$,
where $X$\ depends on the survey presumed\ and $C_{BAO}^{-1}$ is the inverse
covariance matrix \cite{BAO6}.

\begin{widetext}

\begin{table}[H]
\begin{center}
\begin{tabular}{|c|c|c|c|c|c|c|}
\hline
$z_{BAO}$ & $0.106$ & $0.2$ & $0.35$ & $0.44$ & $0.6$ & $0.73$ \\ \hline
$\frac{d_{A}(z_{\ast })}{D_{V}(z_{BAO})}$ & $30.95\pm 1.46$ & $17.55\pm 0.60$
& $10.11\pm 0.37$ & $8.44\pm 0.67$ & $6.69\pm 0.33$ & $5.45\pm 0.31$ \\ 
\hline
\end{tabular}
\caption{Values of $d_{A}(z_{\ast })/D_{V}(z_{BAO})$ for different values of $z_{BAO}$.}
\label{tab2}
\end{center}
\end{table}

\end{widetext}

\begin{itemize}
\item \textbf{Results: }In this part, we have discussed the results obtained
from the statistical MCMC approach with the Bayesian technique. To restrict
the parameters of the model $H_{0}$, $\alpha $, and $\beta $, we use the
combined dataset for the Hubble dataset with 57 data points, the
Pantheon dataset with 1048 sample points, and the BAO dataset with six
points. For this purpose, we employ the total likelihood function and the
chi-square functions, which are given by ${\mathcal{L}}_{T}={\mathcal{L}}%
_{Hubble}\times {\mathcal{L}}_{SNe}\times {\mathcal{L}}_{BAO}$ and $\chi
_{T}^{2}=\chi _{Hubble}^{2}+\chi _{SNe}^{2}+\chi _{BAO}^{2}$. Fig. \ref%
{Contour} shows the $1-\sigma $ and $2-\sigma $ likelihood contours for the
parameters of the model $H_{0}$, $\alpha $, and $\beta $ using the
Hubble+Pantheon+BAO dataset. The best-fit values of the model parameters
estimated are $H_{0}=64.87_{-0.81}^{+0.84}$ $km/s/Mpc$, $\alpha =0.715_{-0.021}^{+0.021}
$, and $\beta =0.1475_{-0.0016}^{+0.0016}$. It is necessary to highlight that
the best-fit value of $H_{0}$ found in this study is close to the value
acquired by the Planck experiment \cite{Planck2020}. We have also found from
Fig. \ref{Contour} that\ the likelihood functions for the
Hubble+Pantheon+BAO dataset are very well matched to a Gaussian
distribution function. Figs. \ref{H} and \ref{Mu} compare our $H(z)$
quadratic expansion model to the widely accepted $\Lambda $CDM model in
cosmology, i.e. $H\left( z\right) =\sqrt{\Omega _{m}^{0}\left( 1+z\right)
^{3}+\Omega ^{\Lambda }}$, for the figure, we choose $\Omega _{m}^{0}=0.315\pm 0.007$ and $H_{0}=67.4\pm 0.5$ $km/s/Mpc$ \cite%
{Planck2020}. The figures also depict the experimental results from Hubble
and Pantheon, with 57 and 1048 data points with errors, respectively, giving
for a clear comparison between the two models.

\begin{widetext}

\begin{figure}[h]
\centering
\includegraphics[scale=0.60]{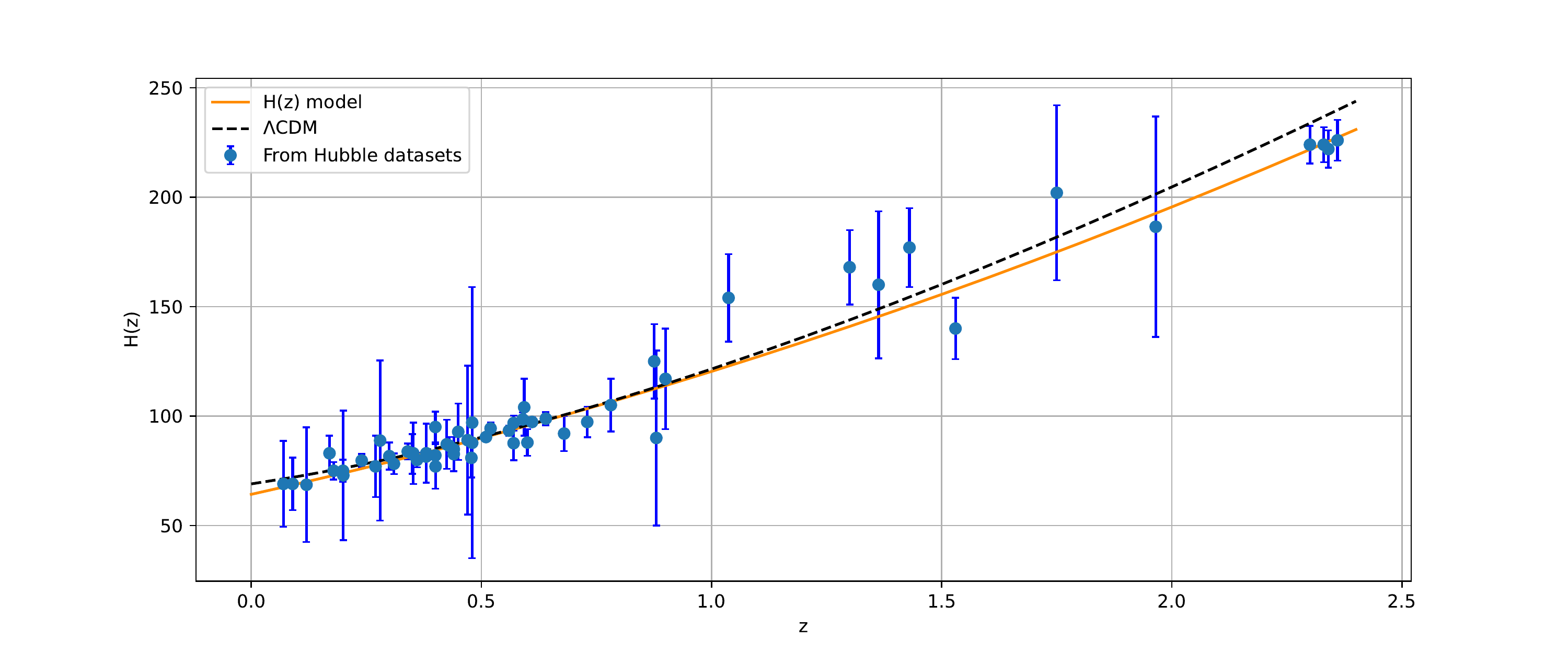}
\caption{The variation of the Hubble parameter $H(z)$ as a function of redshift $z$. The blue dots are error bars, the orange line is the curve produced for our model, and the black dashed line is the standard cosmological model ($\Lambda$CDM).}
\label{H}
\end{figure}	

\begin{figure}[h]
\centering
\includegraphics[scale=0.60]{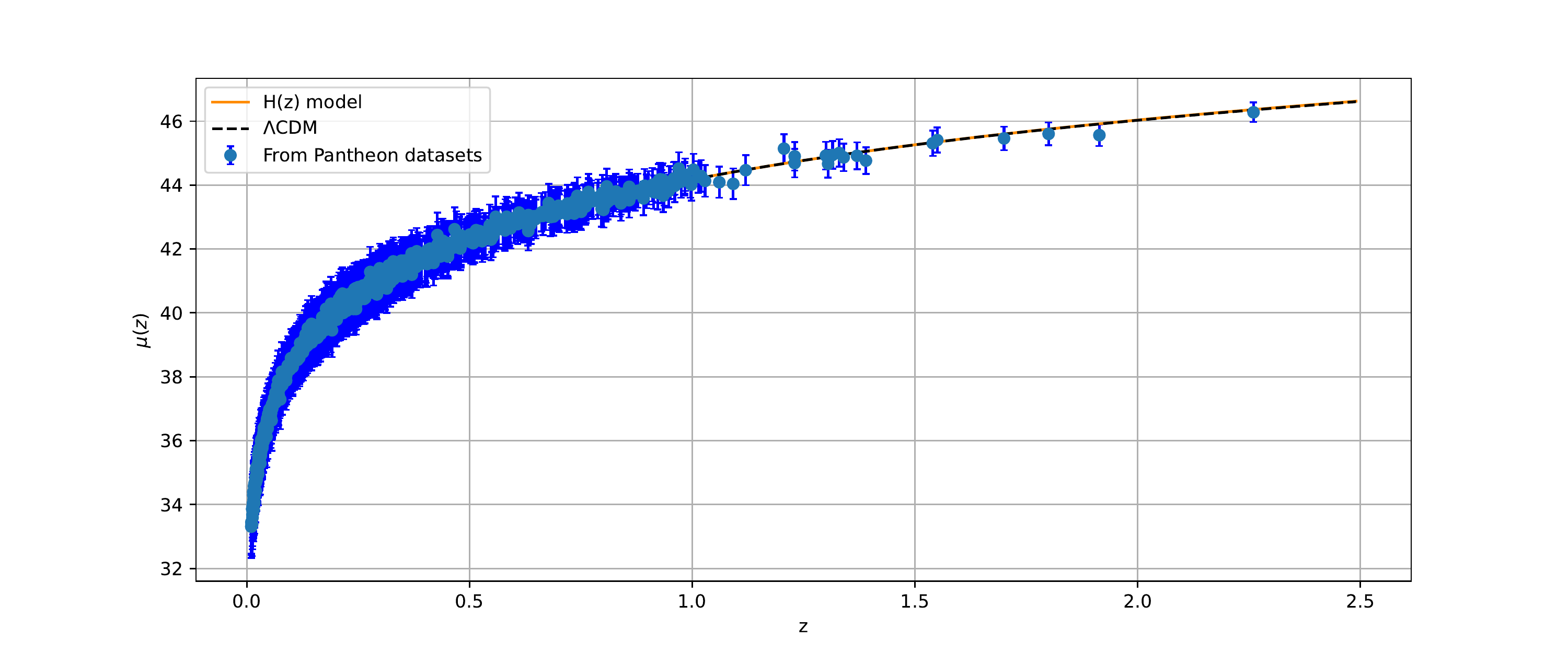}
\caption{The variation of the distance modulus parameter $\mu(z)$ as a function of redshift $z$. The blue dots are error bars, the orange line is the curve produced for our model, and the black dashed line is the standard cosmological model ($\Lambda$CDM).}
\label{Mu}
\end{figure}	

\begin{figure}[h]
\centering
\includegraphics[scale=0.9]{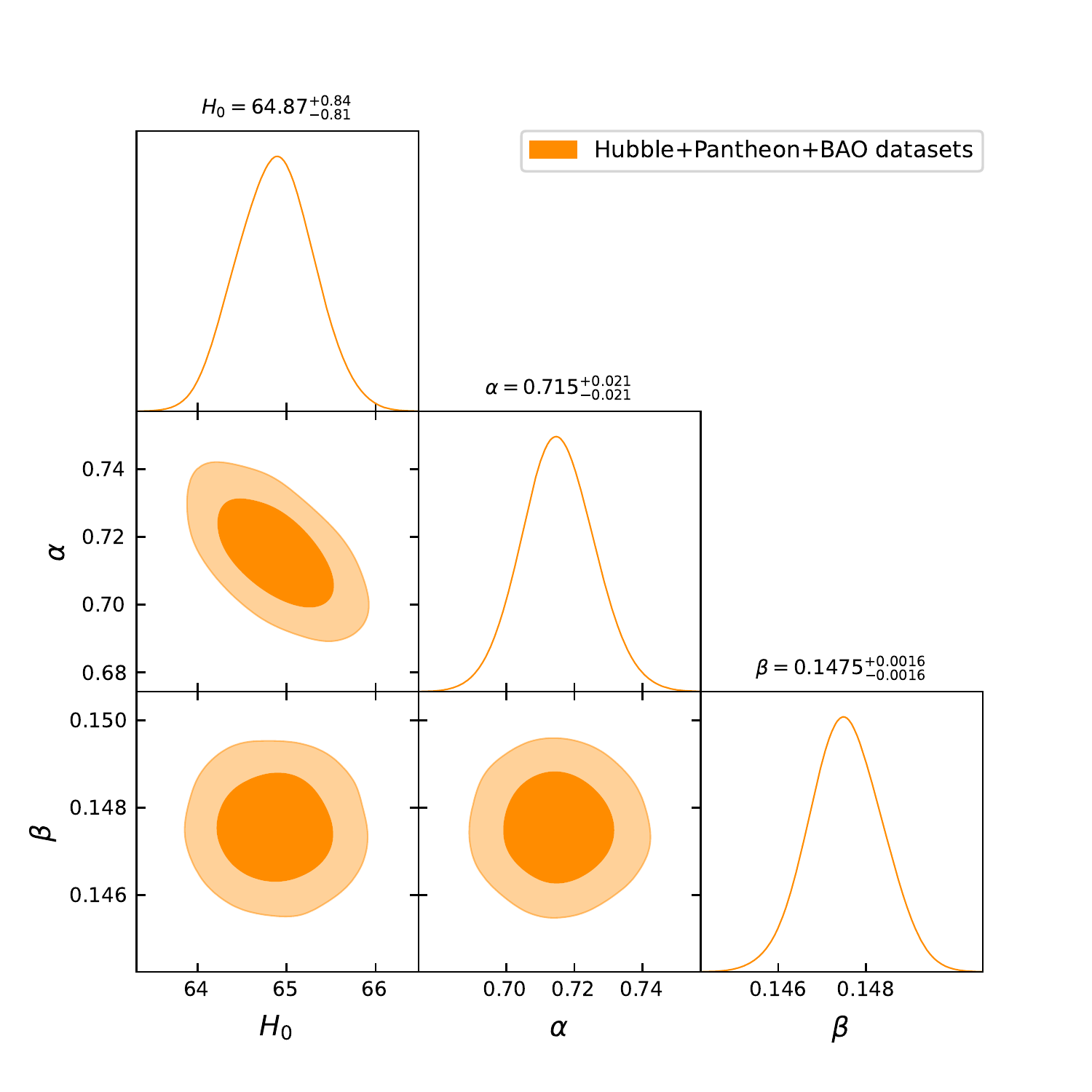}
\caption{The $1-\sigma$ and $2-\sigma$ likelihood contours for the model parameters using the Hubble+Pantheon+BAO dataset. The $1-\sigma$ confidence level (CL) is represented by dark orange shaded regions, while the $2-\sigma$ CL is represented by light orange shaded regions. Also, the parameter constraint values are shown at the $1-\sigma$  CL.}
\label{Contour}
\end{figure}	

\end{widetext}
\end{itemize}

\subsection{Evolution of cosmological parameters}

Here, we will discuss the behavior of cosmological parameters such as the
deceleration parameter, energy density parameter, isotropic pressure, and
EoS parameter. These cosmological parameters are critical for understanding
the evolution and structure of the Universe, as well as testing modified
gravity theories. 

\begin{figure}[h]
\centering
\includegraphics[scale=0.72]{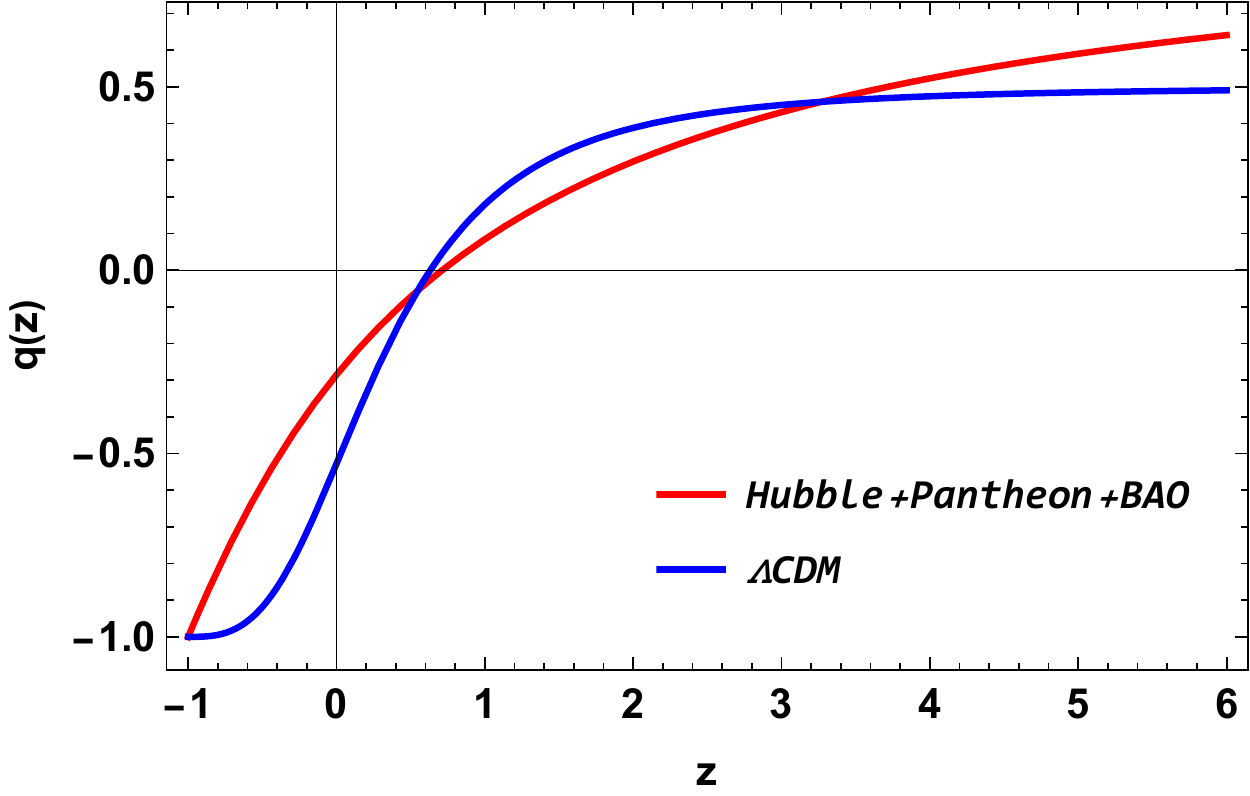}
\caption{The behavior of the deceleration parameter $q$ vs. redshift $z$ using the values constrained from the combined Hubble+Pantheon+BAO dataset. The figure also includes a comparison between our model and the $\Lambda$CDM.}
\label{q}
\end{figure}

According to the observations, the Universe is in a phase transition, which
implies it is transitioning from a past decelerating period to a recent
accelerating expanding period. The physics of this phenomenon may be studied
using a geometrical parameter known as the deceleration parameter $q$, which
explains the Universe's acceleration or deceleration behavior depending on
whether it is negative or positive. If $q>0$, the expansion of the Universe
is decelerating as the gravitational effect of matter and radiation in the
Universe counteracts the expansion. If $q<0$, the Universe's expansion is
accelerating since the repulsive forces of DE surpass the gravitational
effect of matter and radiation. Fig. \ref{q} depicts the behavior of $q$ for
the corresponding values of model parameters restricted by the
Hubble+Pantheon+BAO dataset. It is clear that the Universe accelerates,
decelerates, and shows a phase transition at redshift $z_{t}$. The
expression in Eq. (\ref{qz1}) indicates that $q$ is time-dependent, which
can also represent a Universe phase transition. As $z\rightarrow -1$, we
observe that $q\rightarrow -1$, so, the model exhibits a late time
acceleration. Moreover, for our cosmological model with $H(z)$ quadratic expansion, the constrained value of the deceleration parameter using the combined Hubble+Pantheon+BAO dataset is $q_{0}=-0.285\pm0.021$. This value is in agreement with previous studies that employed similar techniques to estimate $q_{0}$ \cite{DP1,DP2,DP3,DP4}. The transition from deceleration to the acceleration
phase in $f(Q)$ gravity under various assumptions is discussed by \cite{Kou1, Kou2, Kou3, Kou4}. Our
scenario shows clearly that the model is completely in an accelerated phase,
which is consistent with the observed data (see Fig. \ref{q}).

As seen in Fig. \ref{rho}, the density parameter remains positive all
through the Universe's evolution and increases as redshift $z$ increases, as
predicted. It starts as positive and decreases to zero as $z\rightarrow -1$
for various values of $n$ and suitable constraints on $\lambda $. The
isotropic pressure $p$ in Fig. \ref{p} decreases as redshift increases,
starting with a high negative value and tending to a small value at the
current and future epochs for different values of $n$. According to the
observations, the negative pressure is caused by DE in the context
of the Universe's accelerating expansion. As a result, the behavior of
isotropic pressure in our model corresponds to this finding.

\begin{figure}[h]
\centering
\includegraphics[scale=0.70]{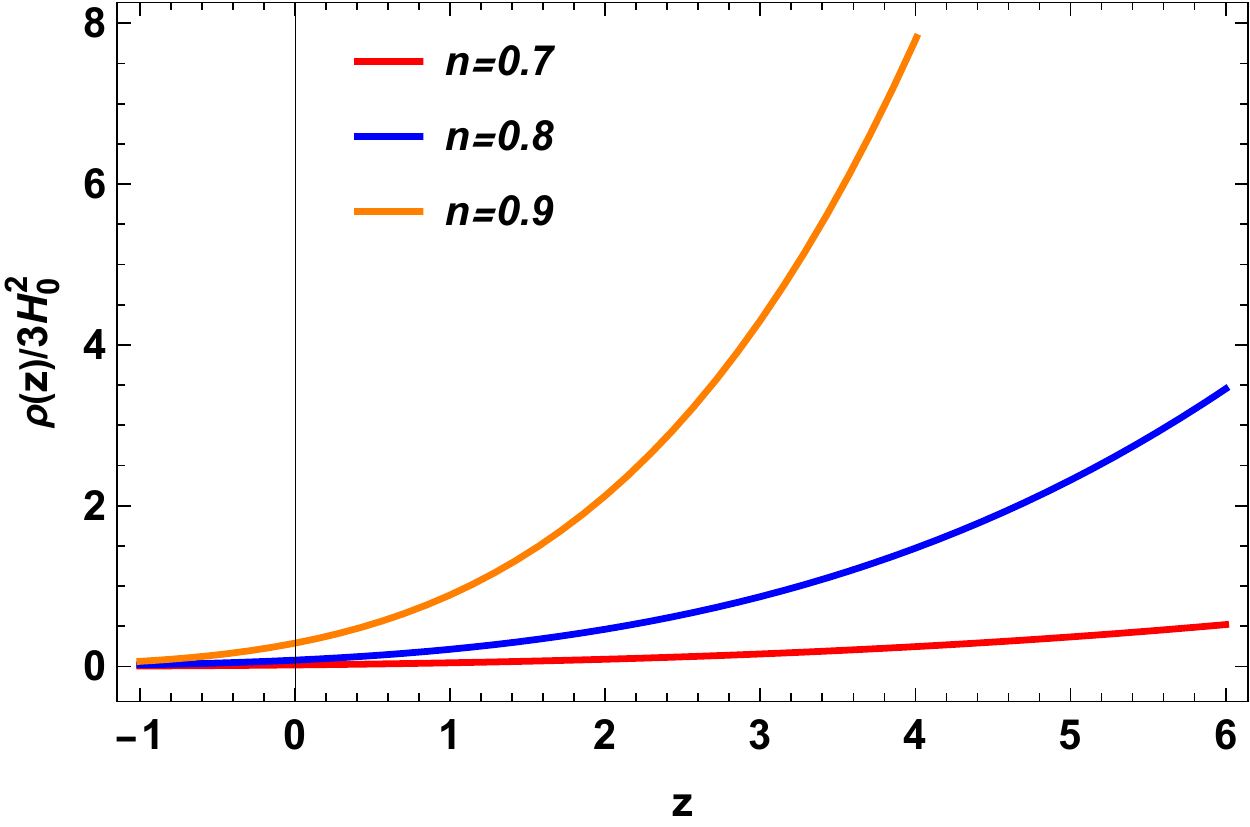}
\caption{The behavior of the density parameter $\rho$ vs. redshift $z$ using the values constrained from the combined Hubble+Pantheon+BAO dataset and and different values of $n$.}
\label{rho}
\end{figure}

\begin{figure}[h]
\centering
\includegraphics[scale=0.70]{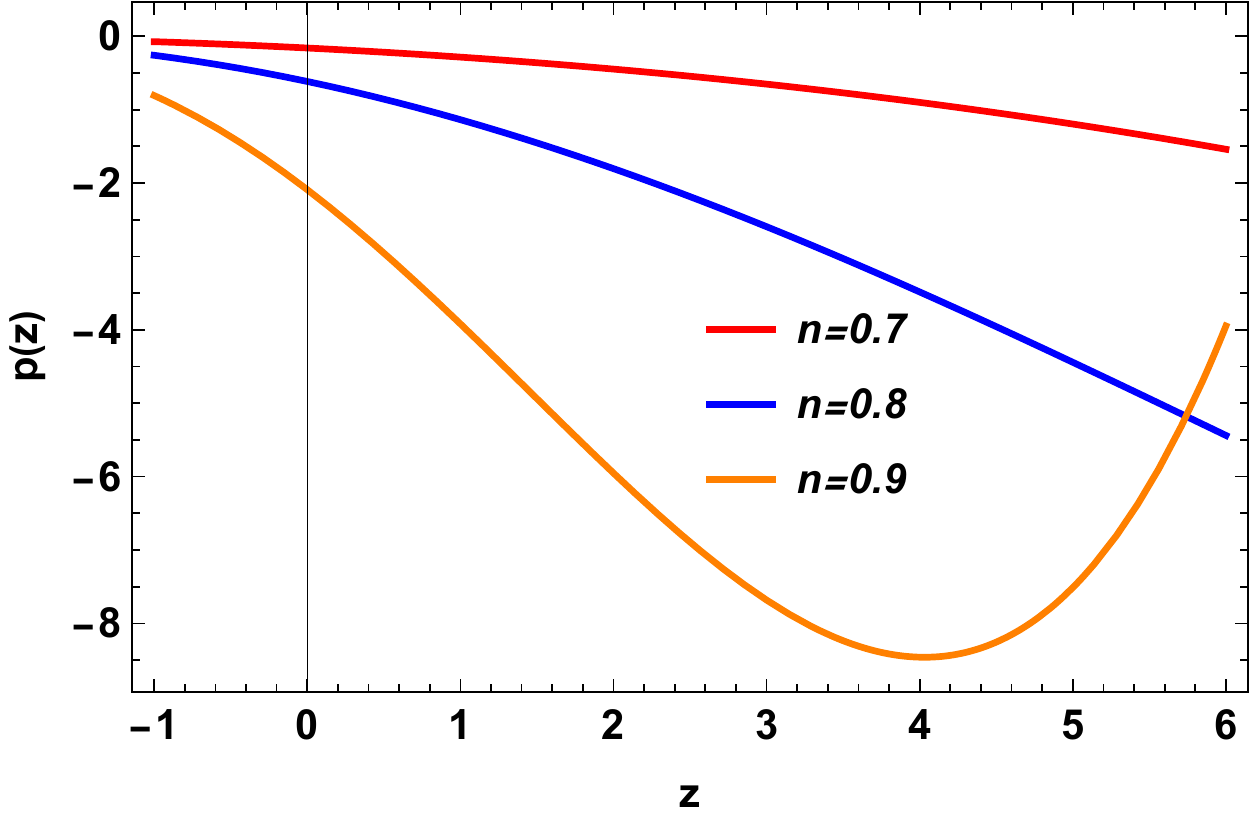}
\caption{The behavior of the pressure $p$ vs. redshift $z$ using the values constrained from the combined Hubble+Pantheon+BAO dataset and and different values of $n$.}
\label{p}
\end{figure}

Recent observational astronomy has estimated a certain range of values for
the EoS parameter $\omega $, which is a function of isotropic pressure and
energy density, i.e. $\omega =\frac{p}{\rho }$, in which the Universe's
expansion scenario is accelerating. The EoS parameter can be useful in
classifying the many periods of accelerated and decelerated Universe
expansion. In the simplest case, the cosmological constant $\Lambda $
emerges for a certain value of the EoS parameter i.e. $\omega =-1$. In
addition, in cosmology, the phantom model and the quintessence model emerge
when $\omega <-1$ and $\omega >-1$, respectively. Fig. \ref{omega} depicts
the evolution of the EoS parameter. It is obvious from Fig. \ref{omega} that 
$\omega <0$ and displays a quintessence DE for various values of $n$%
, indicating an accelerating phase.

\begin{figure}[h]
\centering
\includegraphics[scale=0.72]{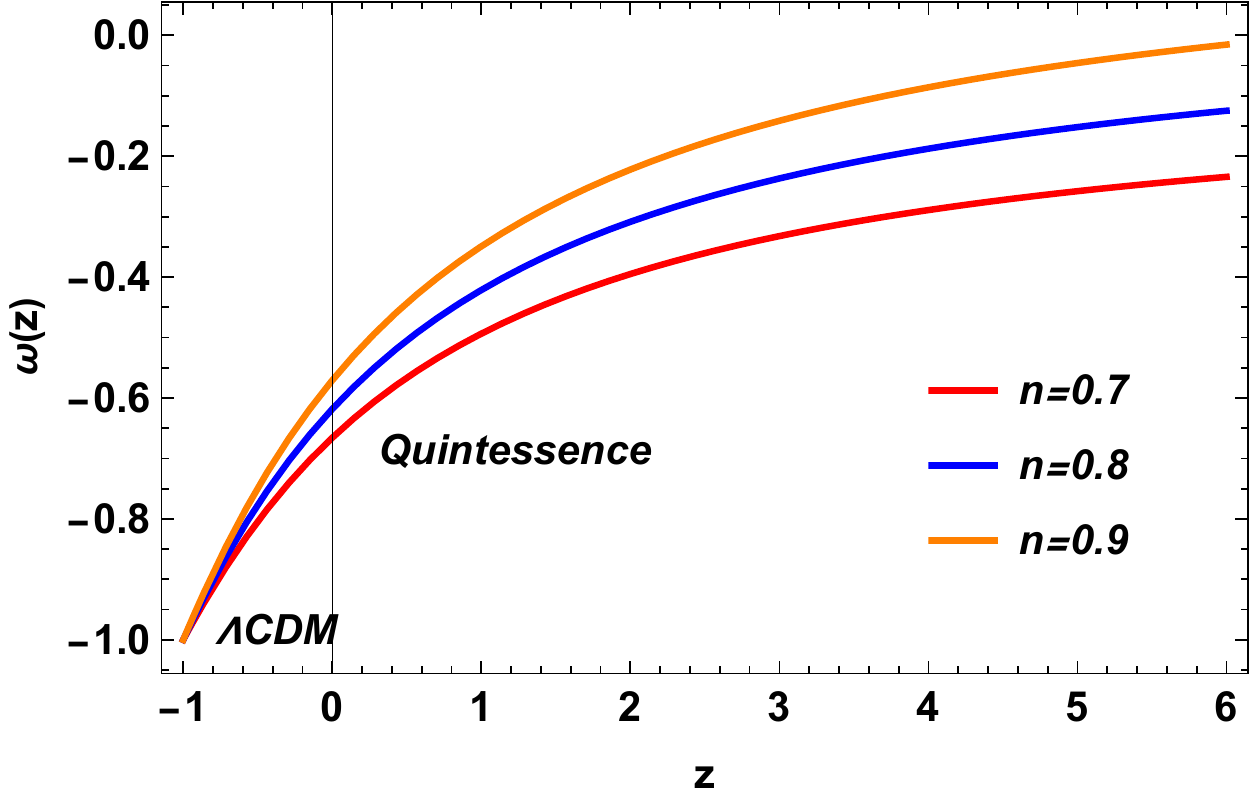}
\caption{The behavior of the EoS parameter $\omega$ vs. redshift $z$ using the values constrained from the combined Hubble+Pantheon+BAO dataset and and different values of $n$.}
\label{omega}
\end{figure}

\section{Stability analysis}

\label{sec4}

In this section, the scalar perturbation analysis will be used to examine
the stability behavior of the $H(z)$ quadratic expansion model in $f(Q)$
gravity. We will follow the linear homogeneous and isotropic perturbation
and discuss the first-order perturbation for the Hubble and density
parameter \cite{Farrugia/2016,Dombriz/2012,Anagnost/2021}. So, the first
order perturbation in the FLRW framework with the perturbation geometry
functions $\delta \left( t\right) $ and matter functions $\delta _{m}\left(
t\right) $ may be represented as,%
\begin{equation}
\widehat{H}(t)=H(t)(1+\delta \left( t\right) )  \label{Hp}
\end{equation}%
\begin{equation}
\widehat{\rho}(t)=\rho (t)(1+\delta _{m}\left( t\right) ),
\end{equation}%
where, $\widehat{H}(t)$, and $\widehat{\rho}(t)$ indicates the perturbed
Hubble and density parameter, $\delta \left( t\right) $ and $\delta
_{m}\left( t\right) $ are the perturbation terms, respectively. As a result,
the perturbation of the functions $f(Q)$ and $f_{Q}$ may be written as  $%
\delta f=f_{Q}\delta Q$ and $\delta f_{Q}=f_{QQ}\delta Q$, where $\delta
Q=12H\delta H$ is the first-order perturbation of the scalar $Q$. Hence,
neglecting the higher power of $\delta \left( t\right) $, the Hubble
parameter may be calculated as $6\widehat{H}^{2}=6H^{2}\left( 1+\delta
\left( t\right) \right) ^{2}=6H^{2}\left( 1+2\delta \left( t\right) \right) $%
. Now, using Eq. \eqref{rho} we obtain%
\begin{equation}
Q\left( f_{Q}+2Qf_{QQ}\right) \delta =-\rho \delta _{m}.
\end{equation}

This yields the matter-geometric perturbation relationship and the perturbed
Hubble parameter may be realized from Eq (\ref{Hp}). Next, to derive the
analytical solution to the perturbation function, consider the perturbation
continuity equation as follows \cite{Agrawal}:
\begin{equation}
\dot{\delta _{m}}+3H(1+\omega )\delta =0.
\end{equation}

Solving the previous equations for $\delta $ and $\delta _{m}$, we derive
the following first order differential equation, 
\begin{equation}
\dot{\delta _{m}}-\frac{3H(1+\omega )\rho }{Q(f_{Q}+2Qf_{QQ})}\delta _{m}=0.
\end{equation}

\begin{figure}[H]
\centering
\includegraphics[scale=0.60]{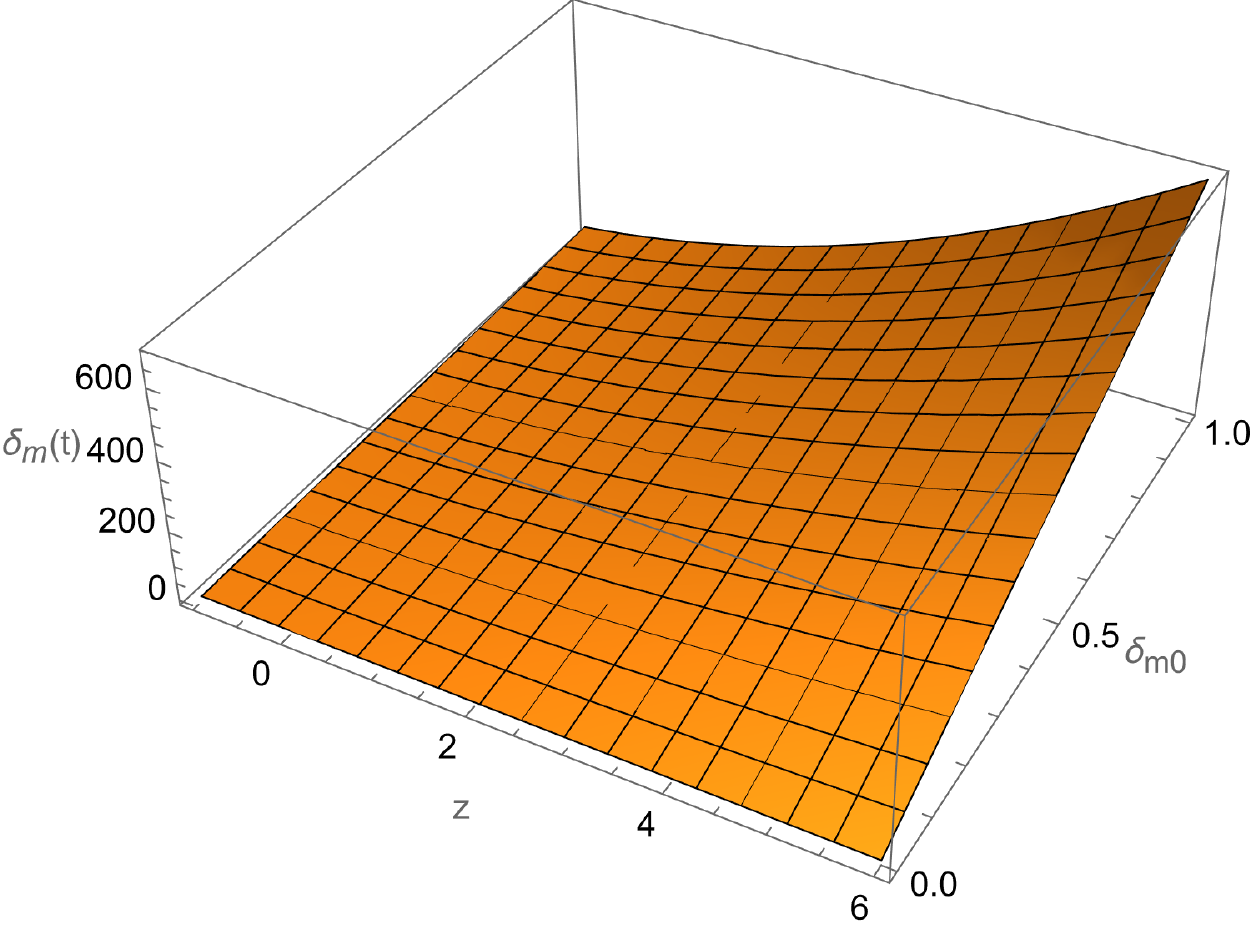}
\caption{The behavior of $\delta _{m}$ vs. redshift $z$ for $n=0.7$ with $0<\delta _{m_{0}}<1$.}
\label{dm1}
\end{figure}

\begin{figure}[H]
\centering
\includegraphics[scale=0.60]{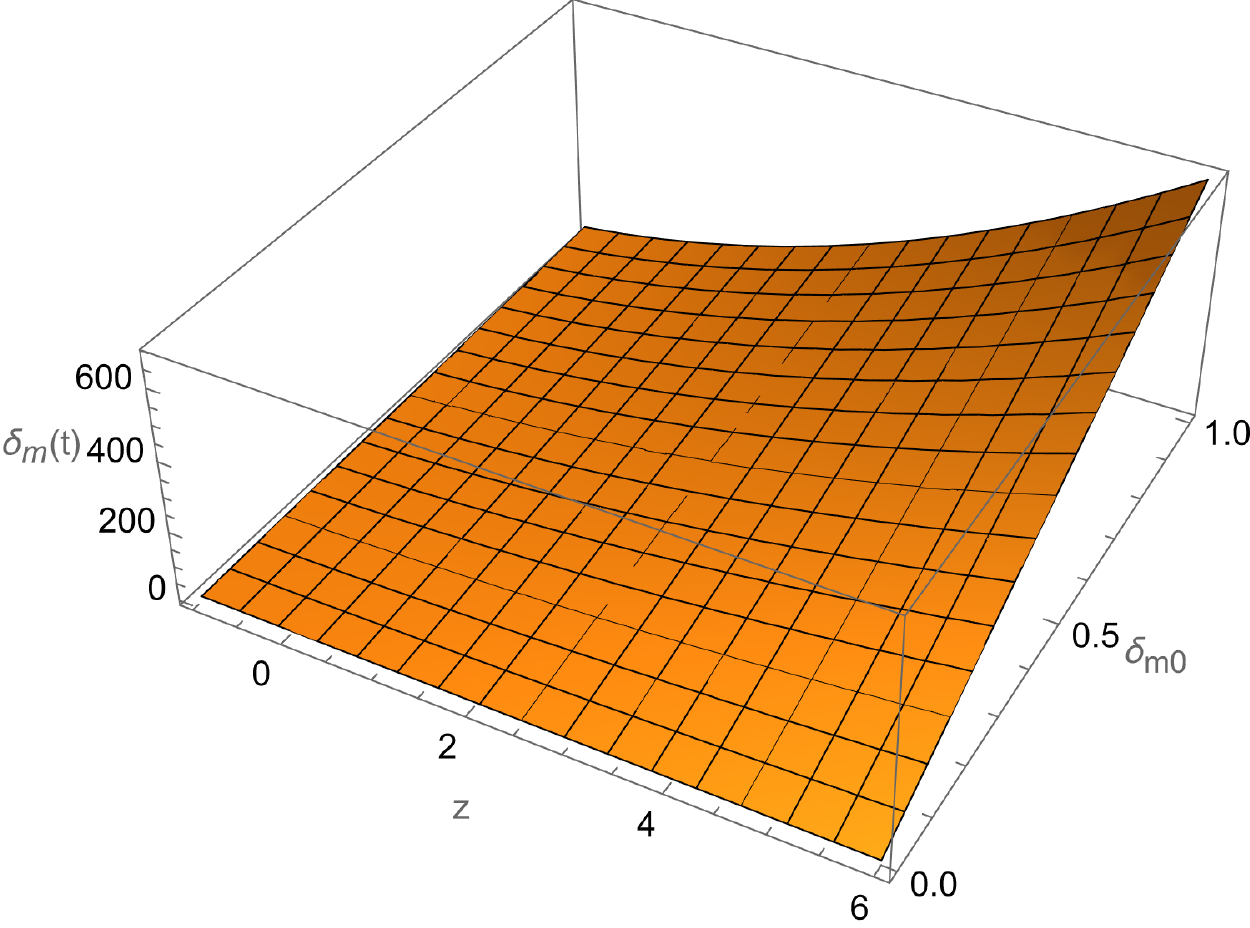}
\caption{The behavior of $\delta _{m}$ vs. redshift $z$ for $n=0.8$ with $0<\delta _{m_{0}}<1$.}
\label{dm2}
\end{figure}

\begin{figure}[H]
\centering
\includegraphics[scale=0.60]{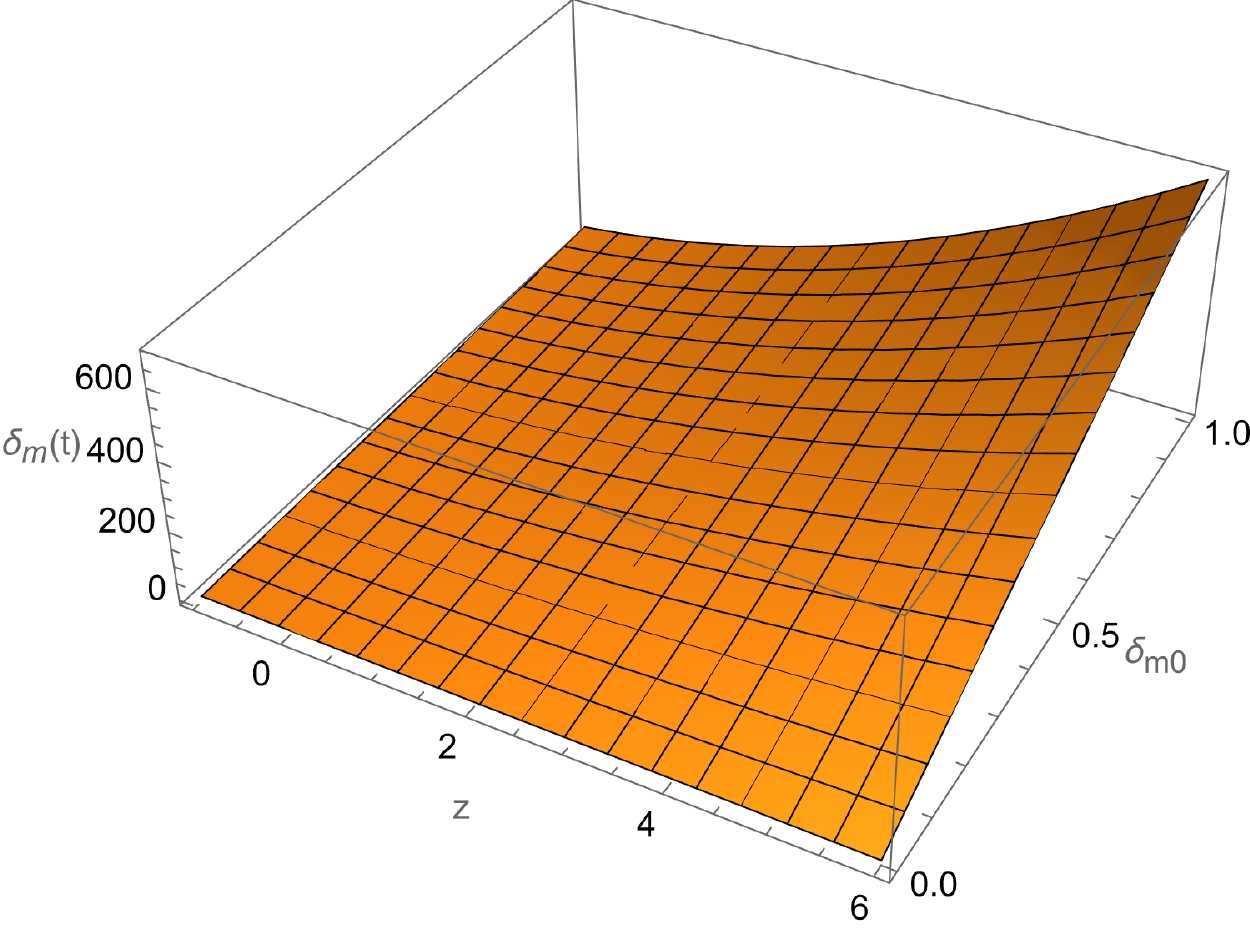}
\caption{The behavior of $\delta _{m}$ vs. redshift $z$ for $n=0.9$ with $0<\delta _{m_{0}}<1$.}
\label{dm3}
\end{figure}

\begin{figure}[H]
\centering
\includegraphics[scale=0.60]{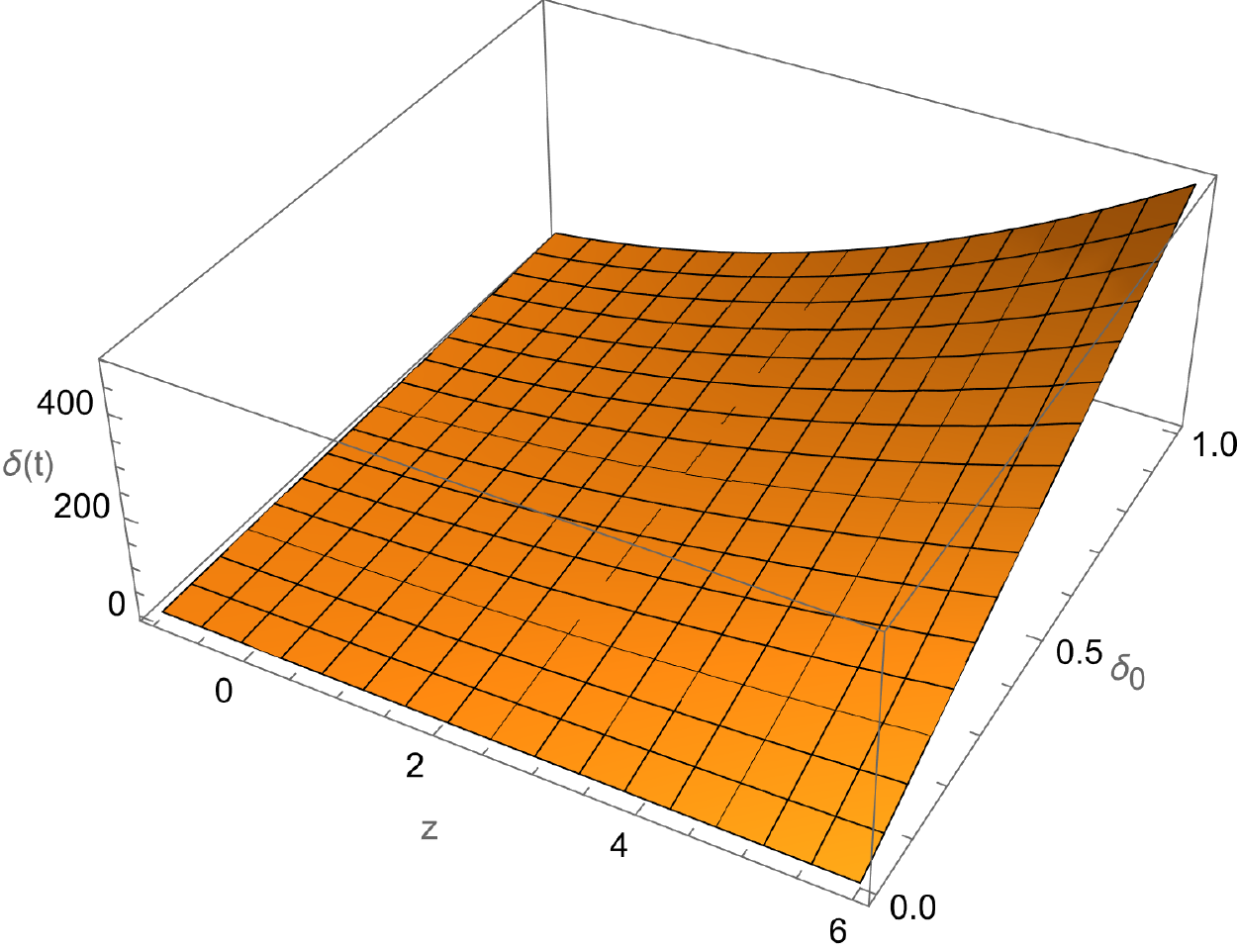}
\caption{The behavior of $\delta$ vs. redshift $z$ for $n=0.7$ with $0<\delta _{0}<1$.}
\label{d1}
\end{figure}

\begin{figure}[H]
\centering
\includegraphics[scale=0.60]{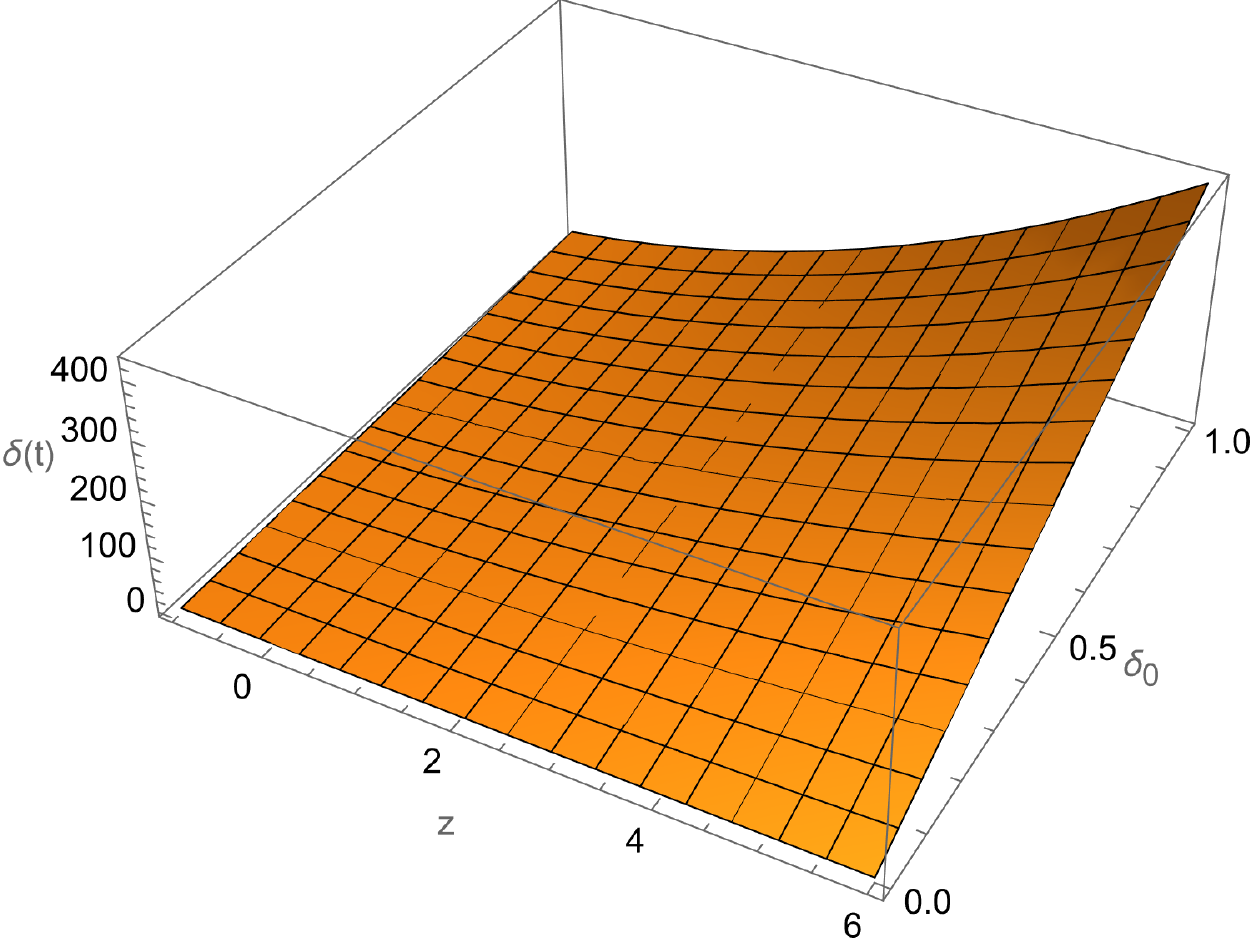}
\caption{The behavior of $\delta$ vs. redshift $z$ for $n=0.8$ with $0<\delta _{0}<1$.}
\label{d2}
\end{figure}

\begin{figure}[H]
\centering
\includegraphics[scale=0.60]{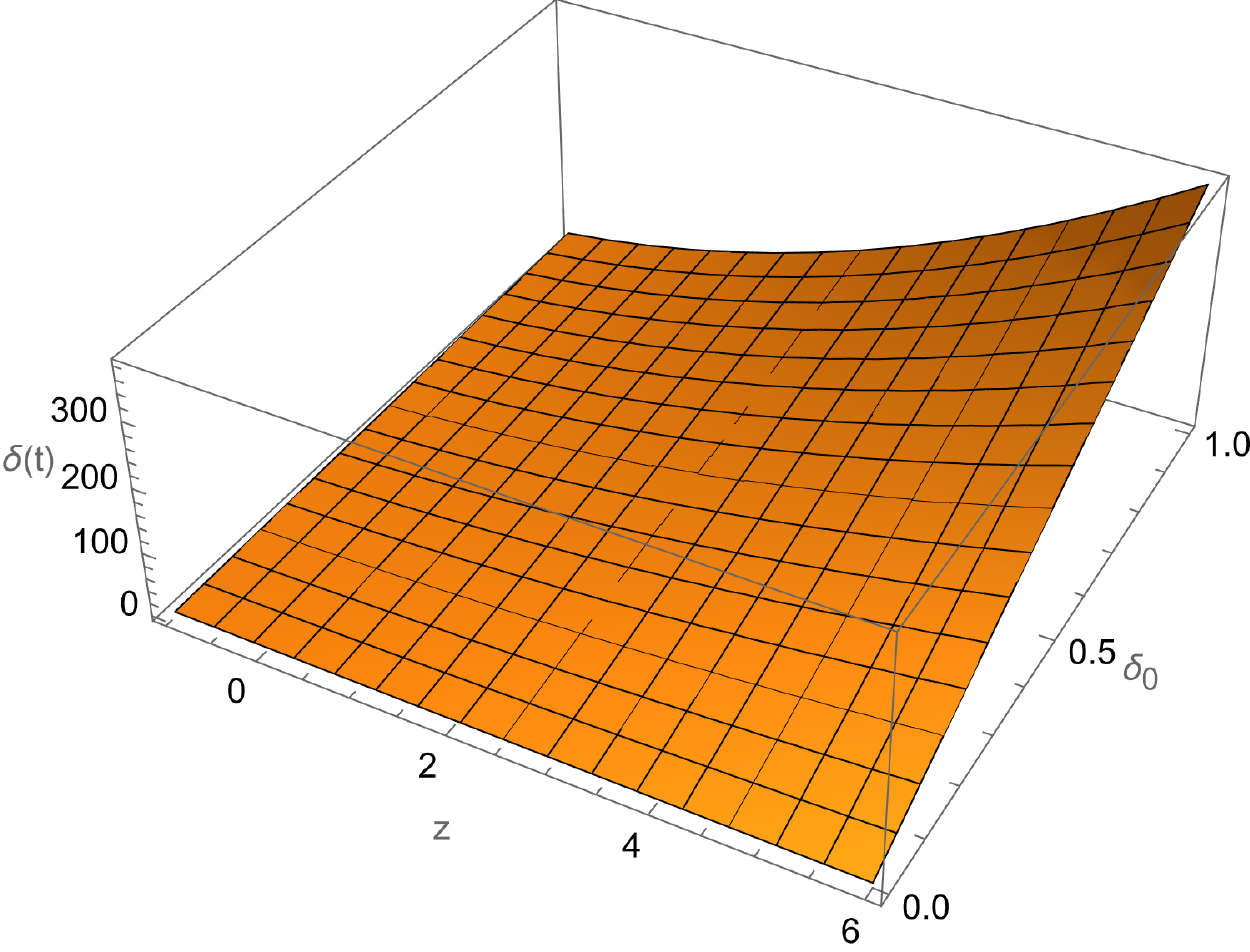}
\caption{The behavior of $\delta$ vs. redshift $z$ for $n=0.9$ with $0<\delta _{0}<1$.}
\label{d3}
\end{figure}

Again, using Eqs. (\ref{rho}) and (\ref{p}) to simplify the above equation,
the solution is written as,%
\begin{eqnarray}
\delta _{m}\left( z\right)  &=&\delta _{m_{0}}H\left( z\right) ,  \notag \\
&=&\delta _{m_{0}}H_{0}\left( 1+\alpha z+\beta z^{2}\right) ,
\end{eqnarray}%
and
\begin{eqnarray}
\delta \left( z\right)  &=&\delta _{0}\frac{\dot{H}}{H},  \notag \\
&=&\delta _{0}(1+z)\frac{dH\left( z\right) }{dz},  \notag \\
&=&\delta _{0}H_{0}(1+z)\left( \alpha +2\beta z\right), 
\end{eqnarray}%
where $\delta _{m_{0}}$ is the integration constant and $\delta _{0}=-\frac{%
\delta _{m_{0}}}{3(1+\omega )}$. \ Fig. \ref{dm1}-\ref{d3} depict the
evolution behavior of perturbation terms $\delta _{m}\left( z\right) $ and $%
\delta \left( z\right) $ in terms of redshift $z$. It is clear that both the
perturbations $\delta _{m}\left( z\right) $ and $\delta \left( z\right) $,
decay quickly and approach zero at late periods. Also, it can be shown that
the behavior of  $\delta _{m}\left( z\right) $ and $\delta \left( z\right) $
is identical for all model parameter values. Thus, our $H(z)$ quadratic
expansion model exhibits stable behavior under the scalar perturbation
method \cite{Agrawal}.

\section{Concluding remarks} \label{sec5}

The present scenario of the Universe's accelerating expansion has gotten increasingly intriguing over time. To develop a proper explanation of the accelerating Universe, several dynamical DE models and modified gravity theories have been used in various ways. In this paper, we have considered a new $f(Q)$ gravity model i.e. $f\left( Q\right) =\lambda _{0}\left(
\lambda +Q\right) ^{n}$ with $H(z)$ quadratic expansion and constrained it with Hubble, Pantheon, and BAO dataset. To constrain the model with observational data, we have used the statistical MCMC approach with the Bayesian method. From our analysis, the best-fit values are found to be $H_0 = 64.87^{+0.84}_{-0.81}$ $km/s/Mpc$, $\alpha = 0.715^{+0.021}_{-0.021}$ and $\beta = 0.1475^{+0.0016}_{-0.0016}$. The fitted value of the Hubble parameter obtained from the study is very close to the value predicted by the Planck experiment. In addition, it is clear that the Hubble quadratic parametrization used in our study is a model-independent way \cite{Mortsell,Cunha,Pacif1} to study the expansion history of the Universe. While the constraints we obtained are not directly related to the parameters of the $f(Q)$ model, they do provide some information about the underlying model when compared to the observational data. Additionally, we performed theoretical analyses to study the behavior of the $f(Q)$ model and its dependence on the parameters, which allowed us to gain insights into the properties of the model and interpret the observational results. We noted that the constraints on $\lambda _{0}$, $\lambda $, and $n$ are obtained indirectly through the Hubble function parametrization, which was chosen to predict the late cosmic acceleration, and are not as stringent as the constraints on $\alpha$ and $\beta$. In the next stage, we investigated the behaviors of some cosmological parameters such as the deceleration parameter, energy
density parameter, isotropic pressure, and EoS parameter. Our results show that the Universe, in this particular model, is in a transition phase. For the higher redshift range, one can see that the Universe is at a decelerated phase with $q(z)>0$ due to the counteraction of matter and radiation gravitational effect on the expansion. For smaller redshift values, the deceleration parameter becomes negative, denoting the Universe's expansion caused by the repulsive forces of DE, which surpass the gravitational effect of matter and radiation. From the EoS parameter behaviour, one can see that the model shows a quintessence behaviour for different values of the model parameter $n$.The model predicts the transition of the Universe from decelerated phase to an accelerated expanding phase. The model predictions are consistent with the observational data. We have also investigated the stability of the model, and it is found that the $H(z)$ quadratic model is stable under the scalar perturbations. 

Finally, the cosmological model investigated in this paper has been constrained, and it is found that it stands in agreement with the  recent observational results. The study of cosmological models in the domain of nonmetricity theory is not very old and recent findings in $f(Q)$ gravity theory show promising aspects in cosmological perspectives. This investigation will contribute to our understanding of $f(Q)$ gravity theory as a promising alternative to GR.

\section*{Acknowledgments}
This research is funded by the Science Committee of the Ministry of Science and Higher Education of the Republic of Kazakhstan (Grant No. AP09058240). M. Koussour is thankful to Dr. Shibesh Kumar Jas Pacif, Centre for Cosmology and Science Popularization, SGT University for some useful discussions. D. J. Gogoi is thankful to Prof. U. D. Goswami, Dibrugarh University for some useful discussions.

\textbf{Data availability} All data used in this study are cited in the references and were obtained from publicly available sources.

\textbf{Conflict of interest} The authors declare that they have no known competing financial interests or personal relationships that could have appeared to influence the work reported in this paper.\newline


\begin{thebibliography}{99}

\bibitem{Planck2020} Planck Collaboration, \textit{Astron. Astrophys.} 
\textbf{641}, A6 (2020).

\bibitem{Starobinsky1} A. A. Starobinsky, \emph{JETP letters} \textbf{86}, 157--163 (2007).

\bibitem{gogoi1} D. J. Gogoi and U. D. Goswami, \textit{Eur. Phys. J. C} \textbf{80}, 1101 (2020) [arXiv:2006.04011].

\bibitem{gogoi2} D. J. Gogoi and U. D. Goswami, \textit{Indian J. Phys.} \textbf{96}, 637 (2022) [arXiv:1901.11277].

\bibitem{gogoi3} D. J. Gogoi and U. D. Goswami, \textit{Physics of the Dark Universe} \textbf{33}, 100860 (2021) [arXiv:2104.13115].

\bibitem{gogoi4}D. J. Gogoi and U. D. Goswami, \textit{J. Cosm. Astropar. Phys.} {\bf 02}, 027 (2023).

\bibitem{Kou1} M. Koussour et al., \textit{Phys. Dark Universe} \textbf{36}, 101051 (2022).

\bibitem{Kou2} M. Koussour et al., \textit{J. High Energy Phys.} \textbf{37}, 15-24 (2023).

\bibitem{Kou3} M. Koussour and M. Bennai, \textit{Chin. J. Phys.} \textbf{79}, 339-347 (2022).

\bibitem{Kou4} M. Koussour et al., \textit{ Ann. Phys.} \textbf{445}, 169092 (2022).

\bibitem{gogoi5}D. J. Gogoi and U. D. Goswami, \textit{J. Cosm. Astropar. Phys.} {\bf 06}, 029 (2022).

\bibitem{Q2} R. Lazkoz, F. S. N. Lobo, M. O. Banos, V. Salzano, \textit{Phys. Rev. D} \textbf{100}, 104027 (2019).

\bibitem{intro3}S. M. Carroll, V. Duvvuri, M. Trodden, and M.S. Turner, \textit{Phys. Rev. D}, \textbf{70}, 043528 (2004).

\bibitem{intro4} S. Nojiri, S. D. Odintsov, and M. Sasaki, \textit{Phys. Rev. D}, \textbf{71}, 123509 (2005). 

\bibitem{intro4a} Nojiri, S., Odintsov, S.D., and
Tretyakov, P.V., \textit{Prog. Theor. Phys. Suppl.} 172, 81 (2008). 

\bibitem{intro4b} Bamba, K., Odintsov, S.D., Sebastiani, L. and
Zerbini, S., [arXiv:0911.4390 [hep-th]].

\bibitem{intro5} Nojiri, S., and Odintsov, S.D 2007. Int. J. Geom. Meth. Mod. Phys., 4, 115. 

\bibitem{intro5a}Cognola, G., Elizalde, E., Nojiri,
S., Odintsov, S.D., and Zerbini, S. 2006, Phys. Rev. D, 73, 084007. 

\bibitem{intro5b}Boehmer, C.G., and Lobo, F.S.N.,
Phys. Rev. D 79, 067504 (2009). 

\bibitem{intro6} A. Habib et al., \textit{Int. J. Mod. Phys. D} 2240015 (2022).

\bibitem{intro7} S. M. Carroll et al., \textit{Phys. Rev. D} \textbf{70},
043528 (2004).

\bibitem{intro7a}D. J. Gogoi and U. D. Goswami, {\it Int. J. Mod. Phys. D} {\bf 31}, 2250048 (2022).
\bibitem{intro8} S. Capozziello et al., \textit{Eur. Phys. J. C} \textbf{80},
2 (2020).

\bibitem{intro9} M. Koussour and M. Bennai, \textit{Class. Quantum Grav}. 
\textbf{39} 105001 (2022).

\bibitem{intro10} Y. F. Cai et al., \textit{Rep. Prog. Phys}. \textbf{79}, 10
(2016).

\bibitem{intro11} J. B. Jimenez et al., \textit{Phys. Rev. D} \textbf{98},
044048 (2018).

\bibitem{intro12} J. B. Jimenez et al., \textit{Phys. Rev. D} \textbf{101},
103507 (2020).

\bibitem{intro15} R. H. Lin and X. H. Zhai, \textit{Phys. Rev. D} \textbf{103}%
, 124001 (2021).

\bibitem{intro16} W. Khyllep, A. Paliathanasis, and J. Dutta, \textit{Phys.
Rev. D} \textbf{103}, 103521 (2021).

\bibitem{intro17} T. Harko et al., \textit{Phys. Rev. D} \textbf{98}, 084043
(2018).

\bibitem{intro18} M. Koussour et al., \textit{J. High Energy Astrophys, }%
\textbf{35, }43-51 (2022).

\bibitem{intro19} Y. Xu et al., \textit{Eur. Phys. J.} \textbf{79}, 8 (2019).

\bibitem{intro20} R. Ferraro and F. Fiorini, \textit{Phys. Rev. D} \textbf{78,}
124019 (2008).

\bibitem{intro21} E. V. Linder, \textit{Phys. Rev. D} \textbf{81}, 127301
(2010).

\bibitem{intro22} C. Q. Geng et al., \textit{Phys. Lett. B} \textbf{704}, 5
(2011).





\bibitem{Jarv} L. Järv et al., \textit{Phys. Rev. D} \textbf{97}, 12 (2018)

\bibitem{Hehl} F. W. Hehl, J. D. McCrea, E. W. Mielke, and Y. Ne’eman,, \textit{Phys. Rept.} \textbf{258}, 1 (1915)

\bibitem{Lazkoz} R. Lazkoz et al., \textit{Phys. Rev. D} \textbf{100}, 104027 (2019).

\bibitem{J1} J. Beltran Jimenez, L. Heisenberg, T. Koivisto, \textit{Phys. Rev. D} \textbf{98}, 044048 (2018).

\bibitem{J2} J. Beltran Jimenez et al., \textit{Phys. Rev. D} \textbf{101}, 103507 (2020).

\bibitem{Mukherjee} A. Mukherjee, and N. Banerjee, N, \textit{Astrophys. Space Sci.} \textbf{352}, 893-898 (2014).

\bibitem{Pacif1} S. K. J. Pacif, \textit{Eur. Phys. J. Plus}, \textbf{135},
10 (2020).

\bibitem{Pacif2} S. K. J. Pacif, R. Myrzakulov and S. Myrzakul, \textit{Int.
J. Geom. Methods Mod.}, \textbf{14}, 07, (2017).

\bibitem{H1} N. Roy, S. Goswami and S. Das, \textit{Phys. Dark Universe}, 
\textbf{36}, 101037 (2022).

\bibitem{H2} J. K. Singh and R. Nagpal, \textit{Eur. Phys. J. C}, \textbf{80}%
, 4 (2020).

\bibitem{DP4} A. Jawad, Z. Khan S. Rani, \textit{Eur. Phys. J. C}, \textbf{80%
}, 1 (2020).

\bibitem{DP5} D. Wang, Y. J. Yan and X. H. Meng, \textit{Eur. Phys. J. C}, 
\textbf{77}, 4 (2017).

\bibitem{DP6} U. Debnath and K. Bamba, \textit{Eur. Phys. J. C}, \textbf{79}%
, 8 (2019).

\bibitem{DP7} A. Jawad et al., \textit{Eur. Phys. J. C}, \textbf{79}, 11
(2019).

\bibitem{qua} J. F. Jesus, R. F. L. Holanda, and S. H. Pereira, \textit{J. Cosmol. Astropart. Phys.} \textbf{2018}, 05 (2018).

\bibitem{Mackey/2013} D. F. Mackey et al., \textit{Publ. Astron. Soc. Pac.} 
\textbf{125}, 306(2013).

\bibitem{Sharov/2018} G.S. Sharov, V.O. Vasilie, \textit{Mathematical
Modelling and Geometry} \textbf{6} 1(2018).

\bibitem{Scolnic/2018} D.M. Scolnic et al., \textit{ApJ} \textbf{859},
101(2018).

\bibitem{BAO1} C. Blake et al., \textit{\ Mon. Not. Roy. Astron. Soc.} 
\textbf{418}, 1707 (2011).

\bibitem{BAO2} W. J. Percival et al., \textit{Mon. Not. Roy. Astron. Soc.} 
\textbf{401}, 2148 (2010).

\bibitem{BAO3} F. Beutler et al., \textit{\ Mon. Not. Roy. Astron. Soc.} 
\textbf{416}, 3017 (2011).

\bibitem{BAO4} N. Jarosik et al., \textit{\ Astrophys. J. Suppl.} \textbf{192%
}, 14 (2011).

\bibitem{BAO5} D. J. Eisenstein et al., \textit{\ Astrophys. J.} \textbf{633}%
, 560 (2005).

\bibitem{BAO6} R. Giostri et al.,\textit{\ J. Cosm. Astropart. Phys.} 
\textbf{1203}, 027 (2012).

\bibitem{DP1} N. Cruz, A. Hernández-Almada, O. Cornejo-Pérez, \textit{Phys. Rev. D}, \textbf{100}, 083524 (2019).

\bibitem{DP2} R. G. Vishwakarma, \textit{Gen. Relativ. Gravit.}, \textbf{33}, 1973-1984 (2001).

\bibitem{DP3} E. González, G. Leon, G. Fernandez-Anaya, \textit{arXiv preprint}, arXiv:2303.16409 (2023).

\bibitem{Farrugia/2016}  G. Farrugia, J. L. Said, Phys. Rev. D, \textbf{94}, 124054 (2016).

\bibitem{Dombriz/2012} A. de la C-Dombriz, D. S-Gomez, Class. Quantum Grav., \textbf{29}, 245014 (2012).

\bibitem{Anagnost/2021} F. K. Anagnostopoulos, S. Basilakos, E. N. Saridakis, Phys. Lett. B, \textbf{822}, 136634 (2021).

\bibitem{Agrawal} A. S. Agrawal, B. Mishra, and P. K. Agrawal , \textit{Eur. Phys. J. C}, \textbf{83}, 2 (2023).

\bibitem{Cunha} J. V. Cunha, J. A. S Lima, \textit{Mon. Notices Royal Astron. Soc.}, \textbf{390}, 210-217 (2008).

\bibitem{Mortsell} E. Mörtsell, C. Clarkson, \textit{J. Cosm. Astropar. Phys.}, \textbf{2009}, 044 (2009).
\end{thebibliography}
\end{document}